\documentclass[superscriptaddress, twocolumn,aps, longbibliography]{revtex4-2}
\usepackage{amsmath, amssymb, color, stmaryrd, wasysym, esint}
\usepackage{scalerel}
\makeatletter
\newsavebox{\@brx}
\newcommand{\llangle}[1][]{\savebox{\@brx}{\(\m@th{#1\langle}\)}%
  \mathopen{\copy\@brx\kern-0.5\wd\@brx\usebox{\@brx}}}
\newcommand{\rrangle}[1][]{\savebox{\@brx}{\(\m@th{#1\rangle}\)}%
  \mathclose{\copy\@brx\kern-0.5\wd\@brx\usebox{\@brx}}}
\makeatother

\makeatletter
\newsavebox{\@brxx}
\newcommand{\lllangle}[1][]{\savebox{\@brxx}{\(\m@th{#1\langle}\)}%
  \mathopen{\copy\@brxx\kern-0.5\wd\@brxx\usebox{\@brxx}\kern-0.5\wd\@brxx\usebox{\@brxx}}}
\newcommand{\rrrangle}[1][]{\savebox{\@brxx}{\(\m@th{#1\rangle}\)}%
  \mathclose{\copy\@brxx\kern-0.5\wd\@brxx\usebox{\@brxx}\kern-0.5\wd\@brxx\usebox{\@brxx}}}
\makeatother

\usepackage{graphicx}% Include figure files
\usepackage{dcolumn}% Align table columns on decimal point
\usepackage{bm}% bold math
\usepackage{xcolor}
\usepackage{xspace}
\usepackage[makeroom]{cancel}
\usepackage{float}
\usepackage{siunitx}
\usepackage{mathtools}
\usepackage{algorithm}
\usepackage{algorithmic}

\definecolor{linkcolor}{rgb}{0,0,0.6} 
\usepackage[pdftex,colorlinks=true,
	pdfstartview = FitV,
	linkcolor    = linkcolor,
	citecolor    = linkcolor,
	urlcolor     = linkcolor,
	hyperindex   = true,
	hyperfigures = false]{hyperref}

\begin{document}
\title{A solid-state theory for dense cylindrical packings of balls}

\author{Luke K. Davis}
\email{luke.davis@ed.ac.uk} 
\affiliation{%
School of Mathematics and Maxwell Institute for Mathematical Sciences, University of Edinburgh, EH9 3FD, Scotland
}%
\affiliation{%
Higgs Centre for Theoretical Physics, University of Edinburgh, EH9 3FD, Scotland
}%

\author{Alexander R. Klotz}
\email{Alex.Klotz@csulb.edu} 
\affiliation{%
Department of Physics and Astronomy, California State University, Long Beach
}%

\begin{abstract}
We develop an analytical theory for the dense packing of hard spheres in cylinders. Physically, our theory consists of a finite cylindrical masking of a close-packed three-dimensional solid and covers the entire range of cylinder aspect ratios, thus going beyond efforts that are focused on very tall cylinders in a narrow range of widths. We explicitly derive an exact equation for resulting packing fractions, valid for any regular lattice, and it provides a basis to understand the oscillations and scaling of volume fractions that have appeared in previous works. Our analytical relation serves as a rigorous lower bound and to tighten it we derive, and implement, an efficient mathematical procedure to optimize the orientation of the cylinder. Furthermore, we suggest simple techniques to improve on the predicted packings. Overall, we provide a general theoretical foundation for the packing of balls in cylinders, valid for all container sizes.
\end{abstract}

\maketitle

\textit{Introduction.}--A classic problem in soft matter is to find the densest packing of balls (hard spheres) into a container, of which a canonical choice is the cylinder \cite{Bogomolov1990,Pickett2000,DurnOlivencia2009,Fu2016,winkelmann2023,Mughal2025,Wang2026}. Insights into cylindrical packing of balls will help to understand other prominent challenges in soft and condensed matter such as the behavior of foams and bubbles \cite{Mann1933,Saadatfar2008,winkelmann2023}, structures of, and within, biological tubules such as viruses, flagella, and microtubules \cite{Erickson1973,Meicenheimer1989}, and the packing inside carbon nanotubes \cite{Troche2004,Sanwaria2014}. Despite resembling the (very difficult) problem of finding the densest packing of spheres in unbounded space \cite{Toth1950,rogers1964packing,Conway1999,Viazovska2017,Cohn2017,Torquato2026}, which has been solved exactly in a few dimensions, understanding dense cylindrical packing of balls has no known exact solution and has attracted significant attention. As such, much work has relied on sophisticated numerical and optimization solutions, \emph{e.g.,} using linear programming and sequential deposition algorithms \cite{Mueller2005,Chan2011,Mughal2012,Fu2016,Fu2017}, with few, if any, schemes to connect these solutions across packing regimes. Thus, there is a clearly missing theory underpinning the physics and mathematics of cylindrical packing of balls.

For cylindrical packing of balls, when a length-scale defining its size, such as the diameter $D = 2R$ or height $H$ of the cylinder, approaches the diameter of the ball, $\sigma$, rich packing geometries and behaviors result \cite{Pickett2000,Mughal2012,Chan2013,Fu2016,Fu2017}. Thus, much of the focus in the field has been exploring cylindrical packing in the narrow regime of $1 \leq D/\sigma  \leq 4$, typically for large heights of the cylinder \cite{Chan2011,Mughal2023}. Considering the other extreme, when all container length-scales are much larger than the ball diameter, $D,H \gg \sigma$, the expectation is that the shape of the container has diminishing influence on the packing, such that the densest packing in the container should approach (for $D/\sigma \rightarrow \infty$) close-packing in unbounded three-dimensional space ($\phi_\text{cp} = \pi/ (3 \sqrt{2})$) \cite{rogers1964packing,Conway1999,Toth1950,Torquato2026}. However, despite detailed knowledge of these regimes, we have no theoretical understanding of the majority of cylindrical packings of balls with size ratios ($D/\sigma \geq 4$).

\begin{figure}[t!]
    \centering
    \includegraphics[width=0.9\linewidth]{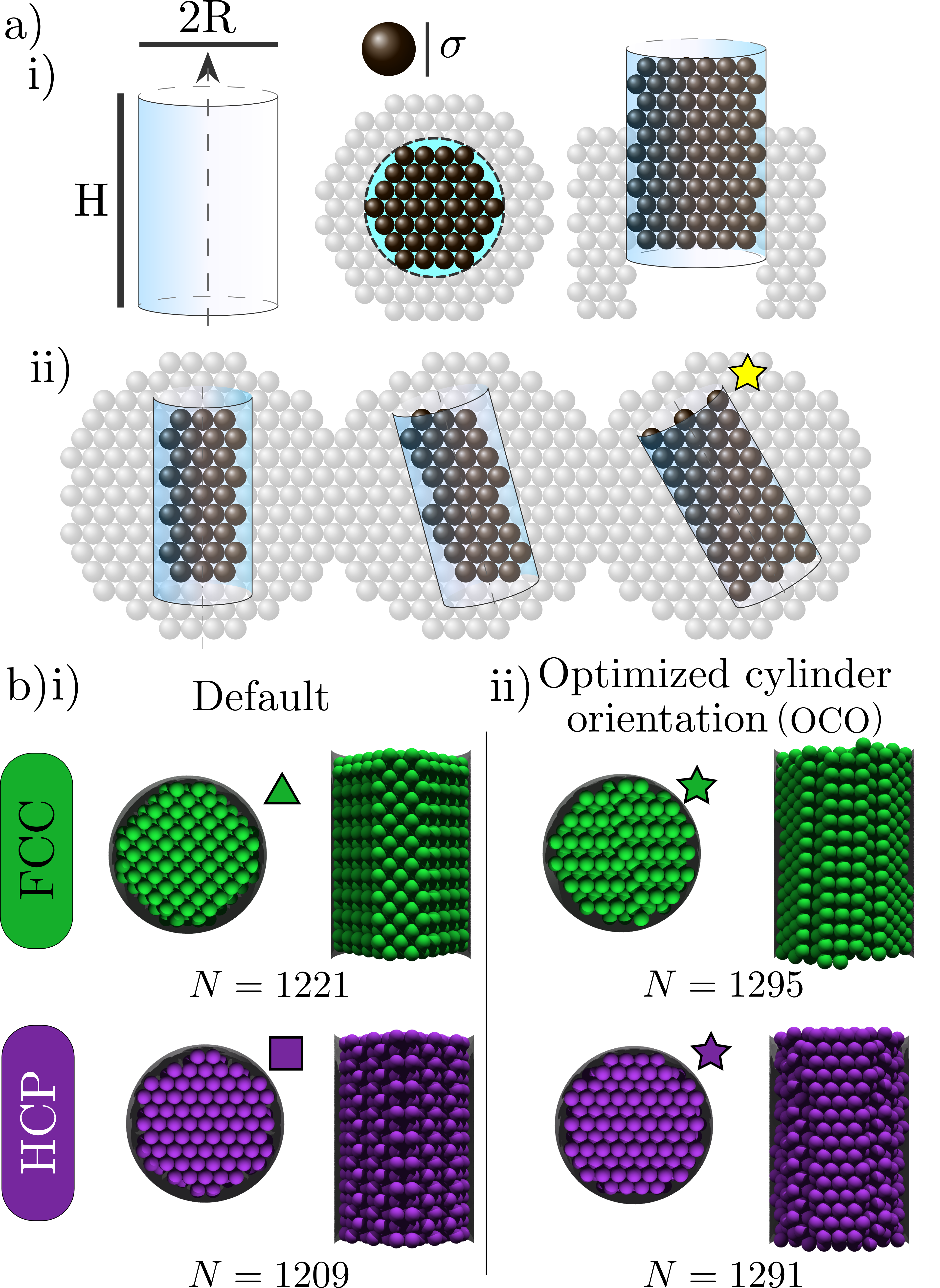}
    \caption{\textbf{Cylindrical packing of balls from masking of a close-packed solid.} a)i)  Generating the default solid-state packings (left) consists of first defining the dimensions of the masking cylinder, its height $H$ and its diameter $2R$, with an upright orientation of [001] (FCC) or [0001] (HCP). ii) A simple optimization of the cylinder orientation improves the packing. b) Examples of default (i) and optimized cylinder orientation (OCO) (ii) cylindrical packings of balls. Cylinder dimensions: $H=30\sigma$ and $2R=10\sigma$.}
    \label{fig:1}
\end{figure}

In this Letter, we present a solid-state theory to obtain an exact lower bound on dense cylindrical packings of balls for the full range of size ratios ($D/\sigma > 1$ and $H/\sigma > 1$). We show that this lower bound tightens for larger containers, and that even for moderate size ratios $D/\sigma \gtrsim 8$ our predicted packings are the best known. We provide a way to extend known narrow-cylinder packings up to $D/\sigma$ between 4 and 8 to provide tighter packings than our lower bound yields. Our theory and calculations are grounded in the foundational physics and mathematics of solids, lattices, and geometry \cite{kittel2004,Ebeling2013}. Like thrusting a paper towel tube through a vat of gumballs, the crux of our approach is the idea of first masking out a finite cylinder of balls from a close-packed solid, whose packing fraction can be solved exactly, and then performing basic (and fast) optimization of the orientation of the cylinder (Fig. \ref{fig:1}).

\textit{Model.}--We consider $N$ monodisperse balls of diameter $\sigma$. The balls are treated as hard spheres where the pairwise distance between any two balls obeys $
|\mathbf{r}_i - \mathbf{r}_j| \geq \sigma$, where $i,j \in \{1, \ldots, N\}$, $i \neq j$, and $\mathbf{r}_i \in \mathbb{R}^3$ is the position of ball $i$. The $N$ balls are forced to lie on a close-packed lattice of a three-dimensional solid, such as a face-centered-cubic (FCC) or hexagonally-close-packed (HCP) lattice. 

\begin{figure*}[t!]
    \centering
    \includegraphics[width=0.95\linewidth]{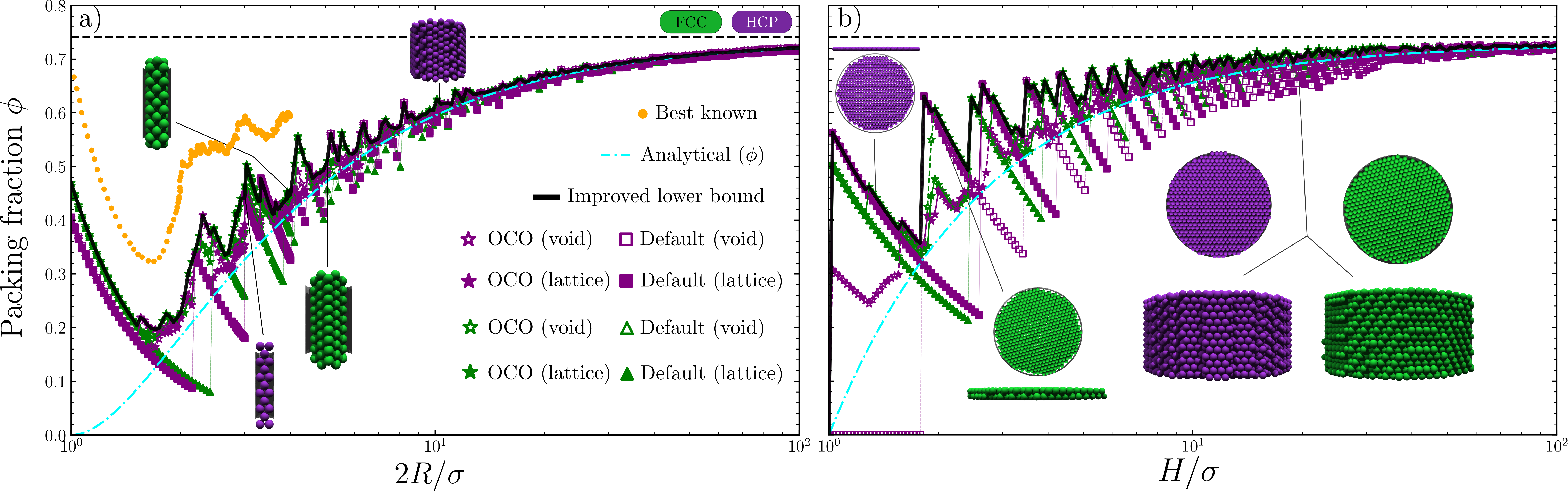}
    \caption{\textbf{Packing fractions for cylindrical packings of balls as predicted by the theory. }\textbf{a)} Packing (volume) fraction for tall ($H/\sigma = 125$) cylinders as a function of the cylinder to sphere diameter ratio. Data points are calculated from the analytically exact reciprocal sum \eqref{eq:PHI_GENERAL_LB}, with the cylinder centered either at the center of a void (void) or coincident with a lattice site (lattice). The cyan curve is the averaged (over all cylinder orientations and positions) analytical packing fraction from Eq. \eqref{eq:PHI_AVG}. The best lower bound is calculated from finding the best packing resulting from all the FCC and HCP optimized cylinder orientation (OCO) packings (Eq. \eqref{eq:extremum_condition}). The best known data are the densest packings found via numerical optimization and are taken from \cite{Mughal2012,Fu2016}. \textbf{b)} Packing fractions for wide ($2R/\sigma=125$) cylinders as a function of cylinder height. The snapshots only show spheres with radial positions within $r < 12.5 \sigma$. }
    \label{fig:2}
\end{figure*}

To obtain the cylindrical geometry of the packing we apply a cylinder mask to the solid. This cylindrical mask, which we call the cylinder for brevity, is located at $\mathbf{R}$ (its midpoint) and has an orientation $\hat{\mathbf{n}}$. We also impose hard boundaries such that, for a cylinder located at the origin, ball positions in the cylinder obey:
\begin{equation}
\begin{aligned}
        x_i^2 + y_i^2 \leq \left(R'\right)^2, \quad 2|z_i| \leq H'
        \label{eq:CylinderBounds},
\end{aligned}
\end{equation}
where $R' \equiv R - \sigma/2$ and $H'\equiv H-\sigma$. The number of balls, $N$, satisfying the above constraints, governs the resulting cylindrical packing fraction $\phi = N \sigma^3/(6R^2 H)$. The densest packing for a given geometry arises from maximizing $N$.

\textit{Theory.}--We now present an explicit analytical solution for the volume fraction of balls, lying on a perfect lattice, that fit inside a cylinder at some prescribed size and orientation. It will be useful to define an axial coordinate about the cylinder center $s(\mathbf{r}):= \hat{\mathbf{n}} \cdot \mathbf{r}$ and a perpendicular distance to its axis as $r_\perp (\mathbf{r}) := |\mathbf{r} - (\hat{\mathbf{n}} \cdot \mathbf{r}) \hat{\mathbf{n}}| = |\mathbf{r} \times \hat{\mathbf{n}}|$. We then define the following indicator function to count spheres in the cylinder:
\begin{equation}
    I(\mathbf{r}) := \Theta\left(R' -\mathbf{r}_\perp(\mathbf{r}) \right) \Theta\left(\frac{H'}{2} - |s(\mathbf{r})| \right), 
\end{equation}
where $\Theta(\cdot)$ is the Heaviside function with the convention $\Theta(0)=1$. We can write the number of balls in terms of the microscopic density $\hat{\rho}(\mathbf{r}) = \sum_{\mathbf{r}' \in \text{lattice}} \delta(\mathbf{r} - \mathbf{r}')$ as:
\begin{equation}
    N = \int d^3 r I(\mathbf{r} - \mathbf{R}) \hat{\rho}(\mathbf{r}).
\end{equation}

For a general crystal lattice, the microscopic density can be written as $\hat{\rho}(\mathbf{r}) = \rho_0 \sum_{\mathbf{G}} S(\mathbf{G}) \exp(i \mathbf{G} \cdot \mathbf{r}),$ where $i =\sqrt{-1}$, $\mathbf{G}$ are the reciprocal lattice vectors, and $S(\mathbf{G})$ is the structure factor of the atomic basis \cite{kittel2004}. This allows us to write the number of balls in a cylinder centered at an arbitrary position $\mathbf{R}$ as:
\begin{equation}
    N(\mathbf{R}) =  \rho_0 \sum_{\mathbf{G}}  S(\mathbf{G}) \exp(i \mathbf{G} \cdot \mathbf{R}) \Tilde{I}(\mathbf{G}),
    \label{eq:NBFourier}
\end{equation}
where $\Tilde{I}(\mathbf{G}) = \int d^3 r I(\mathbf{r}) \exp(i \mathbf{G} \cdot \mathbf{r})$ is the Fourier transform of the indicator function centered at the origin. In End Matter (Sec. \ref{app:DeriveIndicator}) we show that $\Tilde{I}$ can be written as
\begin{equation}
    \Tilde{I}(\mathbf{G}) = 2 \pi (R')^2 H' \frac{J_1(G_\perp R')}{G_\perp R'} \text{sinc} \left( \frac{G_{\parallel} H'}{2} \right),
    \label{eq:Itilde}
\end{equation}
with $J_1(\cdot)$ the Bessel function of the first kind, $\text{sinc}(x) \equiv \sin(x)/x$ (with $\text{sinc}(0)=1$), $G_\parallel = \mathbf{G} \cdot \hat{\mathbf{n}}$, and $G_\perp = |\mathbf{G} \times \hat{\mathbf{n}}|$. Because the density is strictly real and the cylindrical window possesses inversion symmetry ($\Tilde{I}(\mathbf{G}) = \Tilde{I}(-\mathbf{G})$), the imaginary components of the phase sum cancel. For a lattice with a basis of $N'$ points located at relative displacements $\mathbf{d}_j$, we can isolate the real spatial weight $W_{\mathbf{G}}(\mathbf{R})$:
\begin{equation}
    W_{\mathbf{G}}(\mathbf{R}) = \frac{1}{N'} \sum_{j=1}^{N'} \cos\bigl(\mathbf{G} \cdot (\mathbf{R} - \mathbf{d}_j)\bigr).
\end{equation}
Substituting this alongside Eq.~\eqref{eq:Itilde} into Eq.~\eqref{eq:NBFourier}, and factoring out the $\mathbf{G}= \pmb{0}$ term, yields our main result for the continuous packing fraction:
\begin{equation}
    \frac{\phi}{\phi_\text{cp}} =  \frac{V'}{V} \left[1+\sum_{\mathbf{G}\neq \mathbf{0}} W_{\mathbf{G}}(\mathbf{R}) \frac{2J_1(G_\perp R')}{G_\perp R'} \text{sinc} \left( \frac{G_{\parallel} H'}{2} \right) \right],
    \label{eq:PHI_GENERAL_LB}
\end{equation}
where $V' = \pi (R')^2 H'$ and $V = \pi R^2 H$. 

Equation \eqref{eq:PHI_GENERAL_LB} represents an analytically exact lower bound for the packing of any regular crystal lattice in a cylinder for $R/\sigma \geq 1/2$ and $H/\sigma \geq 1$, and is our first main result from the theory. The specific crystal structure only dictates the selection rules for $\mathbf{G}$ and the basis vectors $\mathbf{d}_j$. For a simple Bravais lattice like face-centered cubic (FCC), the primitive basis is a single atom ($\mathbf{d}_1=\mathbf{0}$), yielding $W_{\mathbf{G}}(\mathbf{R}) = \cos(\mathbf{G} \cdot \mathbf{R})$. For hexagonal close-packed (HCP) crystals, the two-atom basis dictates a coupled phase weight.

The above relation, with the appearance of Bessel and trigonometric functions, suggests an origin for the appearance of the sharp peaks and troughs observed in the packing fraction of cylindrical packings of balls \cite{Mughal2012}. The explicit expression for an FCC lattice is shown in the End Matter (Eq. \eqref{eq:PHI_FCC_LB_APP}). Gratifyingly, after performing an average over cylinder positions $\mathbf{R}$, over the unit cell, all the $\mathbf{G} \neq \pmb{0}$ terms vanish, resulting in:
\begin{equation}
       \Bar{\phi}= \phi_{cp} \left(\frac{R}{\sigma}\right)^{-2} \left(\frac{R}{\sigma}-\frac{1}{2}\right)^2 \left(\frac{H}{\sigma}\right)^{-1}\left(\frac{H}{\sigma}-1\right).
       \label{eq:PHI_AVG}
\end{equation}
This equation is our second main result: it is a very simple equation which makes clear predictions for how the packing fraction behaves as a function of the cylinder width and height.

We next sought to produce quantitative packing predictions from the theory (Fig. \ref{fig:2}). To do this, we keep the (upright by default) cylinder center restricted to two points: a lattice site coinciding with sphere centers and the center of the voids (which lie on another lattice \cite{kittel2004}). Our rationale for these simple choices is that they allow the cylinder axis to coincide with many sphere centers, ensuring the inner radial core is the densest. Of course, depending on the type of lattice (FCC or HCP) chosen, and also on the dimensions of the cylinder, one of the choices for its center (a lattice site versus a void) will outperform the other. Further, whilst the reciprocal sum \eqref{eq:PHI_GENERAL_LB} is exact as is, in practice one evaluates it at a finite cutoff 
 $\|\mathbf{G}\| < {G}_\text{max}$. The errors for a sensible (high but not cumbersome) choice of cut-off are shown in the Supplemental Material (see Fig. \ref{fig:2_A1}).
 
First, as expected, we find that the overall trends in the packing fractions predicted from the masked solid (\ref{eq:PHI_GENERAL_LB}), across FCC/HCP and for both varying the width and length of the cylinder, are well captured by the exact averaged packing fraction (\ref{eq:PHI_AVG}) (see Figs. \ref{fig:2}a and b). The predicted solid-state packings are, unsurprisingly, smaller compared to the densest known cylindrical packings of balls \cite{Mughal2012,Fu2016}, shown in the range $1 \leq 2R/\sigma \leq 4$. However, interestingly, some of the character of the best known packing data is borne out from the simple theory, \emph{e.g.,} the large initial drop and then rise in volume fraction for $1 < 2R/\sigma < 3$. As suggested by \eqref{eq:PHI_GENERAL_LB}, the masked-solid packing fractions as a function of cylinder dimensions ($R$ or $H$) have a rich oscillatory (step-like) behavior, though they are most prominent for smaller cylinder dimensions: the amplitude of the oscillations becomes small ($\lessapprox 0.1$) at around $2R/\sigma \gtrsim 10$ (see Fig. \ref{fig:2}a) and $H/\sigma \gtrsim 30$ (Fig. \ref{fig:2}b).

For the full range of cylinder dimensions explored it is not obvious whether there is a superior choice between FCC and HCP (Fig. \ref{fig:2}). However, there are specific cases, across the full range, where we do find that the choice of FCC or HCP is the clearly superior one. Long stretches of superiority occur for relatively small cylinder dimensions. For example, for a very long cylinder, small widths $1 \leq 2R/\sigma \leq 1.1$ tend to favor an FCC lattice (Fig. \ref{fig:2}a). This is mainly due to the fact that the thin cylinder permits only a single column of spheres where, for an upright cylinder, the vertical distance between a single column of spheres is shorter in an FCC (hence denser packing). For a very wide cylinder, short lengths $1 \leq H/\sigma \leq 1.1$ favor HCP with the cylinder centered on a lattice site (Fig. \ref{fig:2}b).

Next, we describe a mathematical optimization procedure to tighten our lower bound (Eq. \eqref{eq:PHI_GENERAL_LB}). The natural degree of freedom to optimize is the cylinder orientation $\hat{\mathbf{n}}$, at a fixed center position. We detail the mathematical procedure for the optimized cylinder orientation (OCO) procedure in Appendix \ref{app:orientation_opt}, however we distill the main recipe here. The OCO is the solution of the optimization problem $\hat{\mathbf{n}}^\ast = \text{argmax}_{\hat{\mathbf{n}}} \phi, \quad || \hat{\mathbf{n}}|| = 1,$ with the maximum of the packing fraction resulting from maximizing the sum in Eq. \eqref{eq:PHI_GENERAL_LB}. Indeed, from making such a sum an objective function, along with a Lagrange multiplier constraint for $ || \hat{\mathbf{n}}|| = 1$, we derive an exact stationarity condition for the optimal cylinder orientation. Ultimately, this leads to solving the following equation
\begin{equation}
\label{eq:extremum_condition}
\hat{\mathbf{n}}^\ast \times \mathbf{A}(\hat{\mathbf{n}}^\ast) = \mathbf{0},
\end{equation}
with $\mathbf{A}(\hat{\mathbf{n}}^\ast)$ given explicitly in the Appendix. The above expression is invariant under reflections, $\hat{\mathbf{n}} \to -\hat{\mathbf{n}}$, and under the cubic point group of the FCC lattice, so in numerical searches one may restrict $\hat{\mathbf{n}}$ to an irreducible wedge of the unit sphere.

Qualitatively, cylindrical packings of balls resulting from the OCO procedure tend to produce denser packing (see Fig. \ref{fig:1}b). Even for cylindrical dimensions that are $\sim 10$-fold larger than the ball radius, the default packings appear to have four-fold (FCC) or six-fold (HCP) rotational symmetry resulting in wasted ``corner'' space. The resulting OCO packings clearly produce more circular (tube-shaped) packings, with the balls filling more space in both the radial and axial directions (Fig. \ref{fig:1}bii).

Quantitatively, our OCO procedure produces improvements on the masked-solid lower bounds \eqref{eq:PHI_GENERAL_LB} (see Fig. \ref{fig:2}). Particularly significant improvements occur for small cylinder dimensions, with diminishing returns for larger cylinders. This is exemplified by the OCO improved lower bound approaching the analytical relation for the packing fraction averaged over cylinder orientation and position. \eqref{eq:PHI_AVG}.  \color{black} However, for very small cylinder dimensions $2R/\sigma, H/\sigma \lesssim 1.1$, despite the OCO packing data being significantly higher than \eqref{eq:PHI_AVG}, the OCO procedure tends to not beat the default, upright, packings. To understand this, we first realize that for these very small cylinder radii or heights, the dimension of the cylindrical packing decreases: for very thin cylinders the effective dimension approaches one, a line, and for very flat (disc-shaped) cylinders the effective dimension approaches two. Then, this reduction in dimension favors cylinder orientations that sit flush with particle centers, which is mainly in the default (upright) orientation. Compared to the default masked-solid packing, the OCO data is closer to the densest packing, known only in the range $1 \leq 2R/\sigma \lessapprox 4$, and is approximately $2/3$ of the best known packing fraction. We expect that for large cylinders, $2R/\sigma > 10$ (for $H/\sigma \gg 1$), the difference between the densest packings and the OCO lower bound will be negligible. Since, to the best of our knowledge, there are no data on the densest packing as a function of $H/\sigma$ (for $2R/\sigma \gg 1$), we do not know (absolutely) how tight the OCO lower bound is for this case. However, qualitatively, for larger heights we expect that the lower bounds predicted by the theory will get closer to the optimal packing fractions.

\begin{figure}[t!]
    \centering
    \includegraphics[width=0.99\linewidth]{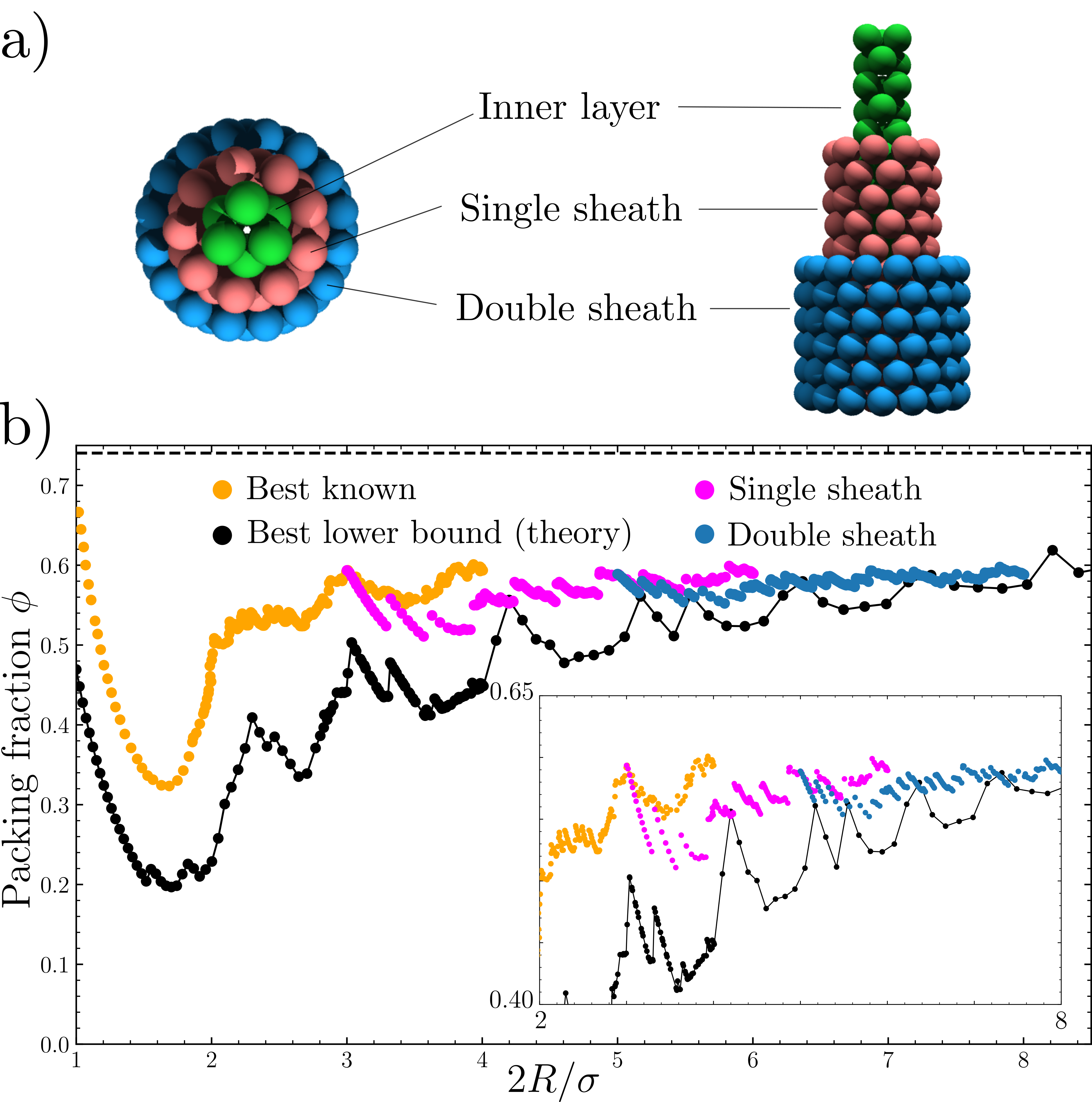}
    \caption{\textbf{The ``sheathing'' procedure: hexagonally packed cylindrical monolayers are added around dense cylindrical packings resulting in highly efficient packings at larger cylinder radii.} \textbf{a)} Visualization of the sheathing technique, shown for an initial packing (the inner layer) at $2R/\sigma$=2.2 with the first sheath at  $2R/\sigma=$4.2, and the second sheath at  $2R/\sigma=$6.2. \textbf{b)} Comparing packing fractions for dense cylindrical packings of balls from the OCO procedure (this work), the sheathing technique (this work), and the best known packings (\cite{Mughal2012,Fu2016}). }
    \label{fig:3}
\end{figure}

After highlighting the range of cylinder diameters ($1 \leq 2R/\sigma \leq 10$) where potentially significant progress can be made, we next wondered whether there were straightforward ways of surpassing the OCO lower bound. To this end, we derived an expression to extend an efficient packing fraction at a given radius $R_0$, using it as a seed to determine a packing at an extended radius $R_E=R_0+\sigma$ by surrounding it with a non-interfering sheath formed from a cylindrical monolayer of hexagonally packed spheres (see Fig. \ref{fig:3}a). This sheathing process may be iterated by adding more concentric layers, and is visualized for two layers. Each monolayer has a volume fraction approaching $\pi/\sqrt{27}\approx0.604$, two-thirds the area fraction of hexagonally packed spheres, which is the fixed point for iterated sheathing. 

If the central configuration is a vertical column of spheres, iterating this sheathing procedure will produce efficient packings at odd integer values of $2R_E/\sigma$. More generally, if a configuration of spheres has packing fraction $\phi_0$ at radius $R_0$, a non-overlapping ring of spheres may be placed at, or on, a circle at radius $R_c=R_0+\sigma/2$. Then, additional rings may be stacked vertically, each layer rotated by half the angle subtended between adjacent spheres such that they form a hexagonal packing as they stack. The number of spheres in each ring arises from the geometric constraints of their non-overlap:

\begin{equation}
    N_s=\bigg\lfloor \frac{\pi}{\arcsin({\sigma/2R_c})} \bigg\rfloor\approx \bigg\lfloor\frac{2\pi R_c}{\sigma}\bigg\rfloor.
\end{equation}

The vertical shift of each ring is typically slightly below $\sqrt{3}\sigma/2$ depending on how closely $R_c$ allows an integer number of spheres to fit into a ring. A vertical stacking of these rings creates a cylindrical monolayer that can be unfolded into a planar hexagonal packing of disks with an area fraction that approaches $\pi/\sqrt{12}$. This sheath can be treated as a hexagonally packed annulus with inner and outer radius $R_0$ and $R_0+\sigma$, thickness $\sigma$, and height $H$, with a volume fraction that is two-thirds the area fraction of its cylindrical cross-section (as a sphere occupies two-thirds of its bounding cylinder). For asymptotically tall cylinders, such that incommensurability between the heights of the inner cylinder and the sheath is not relevant, the volume fraction may be computed as an average of the volume fraction of the inner packing (weighted by $\pi R_0^2$) and the sheath (weighted by $2\pi R_c\sigma$). The volume fraction of the new cylinder of radius $R_E$ formed from the inner seed and the sheath is:

\begin{equation}
\phi'=\frac{1}{R_E^2}\left(\phi_0R_0^2+\frac{N_s\sigma^2}{6\sqrt{1-2\frac{R_c^2}{\sigma^2}\left(1-\cos\frac{\pi}{N_s}\right)}}   \right).   
\end{equation}

This expression is derived from geometry in the Supplemental Material (Eq.~\eqref{eq:phiprime}). By taking the best known packings from $2R/\sigma=1$ to $4$ and extending them with a sheath, we can find packings that are more efficient than the confined OCO solution (with a small number of exceptions) from 3 up to 6 (Fig. \ref{fig:3}b). Extending those solutions recursively with a second sheath yields good configurations up to $2R/\sigma=8$, beyond which the OCO, and eventually Eq.~\eqref{eq:PHI_AVG}, solution is typically superior. To the extent that higher packing fractions can be discovered, such effort is only likely to be worthwhile in the narrow range of $2R/\sigma$ from 4 to 6. Discovering the densest packing for a very wide cylinder, at varying height, (Fig. \ref{fig:2}b) is largely open.

\textit{Conclusions.}--We have established an analytical solid-state theory for the packing of hard spheres in cylindrical confinement, bridging the long-standing gap between packing in the thin-cylinder limit and unbounded three-dimensional close-packing. Our approach rests on mathematically masking perfect FCC and HCP lattices, where we derived an exact analytical lower bound for the packing fraction valid across all cylinder widths and heights. This formulation provides a basis for the oscillatory behavior observed in the packing fraction in cylindrically confined packings and, by averaging over cylinder orientations and placements, we arrived at a compact equation which predicts overall trends. Furthermore, our optimized cylinder orientation (OCO) procedure systematically tightened this bound, generating the densest known analytical packings for wider cylinders. For intermediate regimes, our sheathing construction provides an accessible route to denser packing without relying on laborious numerical optimization. Ultimately, this framework dispenses with the need for exhaustive numerical searches for large cylinders, providing a rigorous, physically grounded program which opens the door to deepening our understanding of sphere packing in other container geometries, polydisperse systems, and higher spatial dimensions.

\textit{Acknowledgments.}--L.K.D. acknowledges funding from the Flora Philip Fellowship at the University of Edinburgh. A.K. acknowledges funding from the National Science Foundation, grant number 2336744. This work resulted from inspiration garnered at the International Conference on Packing Problems (ICPP) 2026.

\textit{Data availability.}--The data supporting this study will be made openly available upon publication. 

\textit{AI Disclosure.}--Gemini (v.3.1 Pro) was used to generate first-attempt code for the (tachyon) rendering of already generated cylindrical packings of balls. The output was verified through visual inspection. 

\twocolumngrid
\bibliography{refs}

\onecolumngrid
\begin{center}
    \textbf{END MATTER}
\end{center}
\twocolumngrid
\section{Deriving the indicator function}
\label{app:DeriveIndicator}

To evaluate the Fourier transform of the cylindrical indicator function, we begin with its formal definition over the spatial domain:
\begin{equation}
    \tilde{I}(\mathbf{G}) = \int d^3r\, I(\mathbf{r}) e^{i \mathbf{G} \cdot \mathbf{r}}.
\end{equation}
The indicator function $I(\mathbf{r})$ restricts the integration volume to the interior of a cylinder of radius $R'$ and height $H'$, whose principal axis is oriented along the unit vector $\hat{\mathbf{n}}$. 

Given the cylindrical geometry, it is natural to shift to a cylindrical coordinate system $(r_\perp, \theta, z)$. We define the axial coordinate as $z = \mathbf{r} \cdot \hat{\mathbf{n}}$ and the transverse radial vector as $\mathbf{r}_\perp = \mathbf{r} - z\hat{\mathbf{n}}$. The spatial volume measure transforms accordingly as $d^3r = r_\perp \, dr_\perp \, d\theta \, dz$.

We proceed by decomposing the reciprocal lattice vector $\mathbf{G}$ into its components parallel and perpendicular to the symmetry axis:
\begin{align}
    G_\parallel &= \mathbf{G} \cdot \hat{\mathbf{n}}, \\
    \mathbf{G}_\perp &= \mathbf{G} - G_\parallel \hat{\mathbf{n}}.
\end{align}
The phase factor in the integrand then separates additively, such that $\mathbf{G} \cdot \mathbf{r} = G_\parallel z + \mathbf{G}_\perp \cdot \mathbf{r}_\perp$. By defining $\theta$ as the relative azimuthal angle between $\mathbf{G}_\perp$ and $\mathbf{r}_\perp$, the transverse dot product reduces to $\mathbf{G}_\perp \cdot \mathbf{r}_\perp = G_\perp r_\perp \cos\theta$.

This decomposition allows the volume integral to factorize exactly into a one-dimensional axial integral and a two-dimensional planar integral over the cross-section, which reads as
\begin{equation}
    \tilde{I}(\mathbf{G}) = \left( \int_{-H'/2}^{H'/2} dz\, e^{i G_\parallel z} \right) \left( \int_{0}^{R'} dr_\perp\, r_\perp \int_{0}^{2\pi} d\theta\, e^{i G_\perp r_\perp \cos\theta} \right).
\end{equation}

Evaluating the axial integral is straightforward and yields the $\mathrm{sinc}(\cdot)$ function:
\begin{equation}
\begin{aligned}
    \int_{-H'/2}^{H'/2} dz\, e^{i G_\parallel z} &= \frac{e^{i G_\parallel H'/2} - e^{-i G_\parallel H'/2}}{i G_\parallel} \\
    &= H' \, \mathrm{sinc}\left(\frac{G_\parallel H'}{2}\right),
    \end{aligned}
\end{equation}
where we have employed the unnormalized definition $\mathrm{sinc}(x) \equiv \sin(x)/x$.

For the planar integral, we first integrate out the azimuthal degree of freedom. We recognize this angular integral as the standard integral representation of the zeroth-order Bessel function of the first kind:
\begin{equation}
    \int_{0}^{2\pi} d\theta\, e^{i G_\perp r_\perp \cos\theta} = 2\pi J_0(G_\perp r_\perp).
\end{equation}

Next, we evaluate the remaining radial integral. Substituting the angular result back in, we make the dimensionless change of variables $u = G_\perp r_\perp$, which gives $du = G_\perp dr_\perp$. Utilizing the well-known Bessel function identity $\int u J_0(u) du = u J_1(u)$, we find:
\begin{equation}
\begin{split}
    2\pi \int_{0}^{R'} dr_\perp\, r_\perp J_0(G_\perp r_\perp) 
    &= \frac{2\pi}{G_\perp^2} \int_{0}^{G_\perp R'} du\, u J_0(u) \\
    &= \frac{2\pi}{G_\perp^2} \Big[ u J_1(u) \Big]_0^{G_\perp R'} \\
    &= 2\pi (R')^2 \frac{J_1(G_\perp R')}{G_\perp R'}.
\end{split}
\end{equation}

Finally, taking the product of the axial and planar spatial contributions recovers the exact analytical form for the Fourier transform of the cylindrical volume:
\begin{equation}
    \tilde{I}(\mathbf{G}) = 2\pi (R')^2 H' \frac{J_1(G_\perp R')}{G_\perp R'} \mathrm{sinc}\left(\frac{G_\parallel H'}{2}\right).
\end{equation}
This completes the derivation of Eq.~\eqref{eq:Itilde}.

\section{Explicit form of the packing fraction}
\label{app:ExplicitPhiFCC}

For a close-packed lattice (FCC or HCP), note that $\rho_0 v_{d=3}(\sigma) = \phi_{cp} = \pi/(3 \sqrt{2})$. The full lower bound for the packing fraction thus reads as:
\begin{widetext}
    \begin{equation}
\begin{aligned}
\phi(R,H,\mathbf{R},\hat{\mathbf{n}}) =& \phi_{\mathrm{cp}} \frac{V'}{V} \Biggl[ 1 + \sum_{h,k,l}
\cos(\mathbf{G}_{hkl}\cdot\mathbf{R}) \frac{2J_1(G_{\perp,hkl} R')}{G_{\perp,hkl} R'}
\mathrm{sinc}\left(\frac{G_{\parallel,hkl} H'}{2}\right) \Biggr],
\label{eq:PHI_FCC_LB_APP}
\end{aligned}
\end{equation}
\end{widetext}
and,
\begin{align}
\label{eq:G_hkl}
\mathbf{G}_{hkl} &= \frac{2\pi}{a}\,(h,k,l), \quad \text{for } h,k,l \in \mathbb{Z} \text{ (all even or all odd)}, \\
G_{\parallel,hkl} &= \mathbf{G}_{hkl}\cdot\hat{\mathbf{n}}, \label{eq:G_parallel} \\
G_{\perp,hkl} &= \bigl\| \mathbf{G}_{hkl} \times \hat{\mathbf{n}} \bigr\| 
= \sqrt{\|\mathbf{G}_{hkl}\|^2 - G_{\parallel,hkl}^2}. \label{eq:G_perp}
\end{align}

\clearpage
\appendix
\onecolumngrid
\renewcommand\thefigure{\thesection.\arabic{figure}} 
 \setcounter{figure}{0}

\begin{center}
    \Large Supplemental material
\end{center}

%=========================================================
\section{Optimized cylinder orientation (OCO) procedure}
\label{app:orientation_opt}
%=========================================================

For the following we assume an FCC lattice, however it is straightforward to work with another Bravais lattice such as hexagonal close-packing (HCP) lattice. Because the cylinder is centered on a lattice site, the translational phase factor becomes 
$\cos(\mathbf{G}_{hkl} \cdot \mathbf{R}) = 1$. The resulting packing fraction can be written as
\begin{equation}
\label{eq:phi_RHn}
\begin{split}
\phi(R,H;\hat{\mathbf{n}}) &=  \phi_{\mathrm{cp}}\, \frac{V'}{V} \times \Biggl[ 1 + \sum_{h,k,l}^{\prime} \frac{2J_1(G_{\perp,hkl}R')}{G_{\perp,hkl}R'}\, \mathrm{sinc} \left(\frac{G_{\parallel,hkl}H'}{2}\right) \Biggr],
\end{split}
\end{equation}
where $\sum'$ indicates the FCC selection rule (all-even or all-odd indices) above and excludes $(h,k,l)=(0,0,0)$.

Since $R, H, \sigma$ are fixed in what follows, maximising $\phi$ over $\hat{\mathbf{n}}$ 
is equivalent to maximising the bracketed orientation-dependent factor
\begin{equation}
\label{eq:F_orientation}
\mathcal{F}(\hat{\mathbf{n}}) = 1 + \sum_{h,k,l}^{\prime} f(u_{hkl})\, s(v_{hkl}),
\end{equation}
where
\begin{equation}
\label{eq:uv_defs}
u_{hkl} = G_{\perp,hkl}R', \qquad v_{hkl} = \frac{G_{\parallel,hkl}H'}{2},
\end{equation}
with the functions defined as
\begin{equation}
\label{eq:fs_defs}
f(u) = \frac{2J_1(u)}{u}, \qquad s(v) = \mathrm{sinc}(v) = \frac{\sin v}{v}.
\end{equation}
We then define the orientation-optimised packing fraction as
\begin{equation}
\label{eq:phi_max}
\begin{split}
\phi_{\max}(R,H) &= \max_{\|\hat{\mathbf{n}}\|=1}\; \phi(R,H;\hat{\mathbf{n}}) = \Theta \left(R-\frac{\sigma}{2}\right) \Theta(H-\sigma)\, \phi_{\mathrm{cp}}\, \frac{V'}{V} \max_{\|\hat{\mathbf{n}}\|=1}\; \mathcal{F}(\hat{\mathbf{n}}).
\end{split}
\end{equation}

\subsubsection*{Stationarity condition for the optimal axis $\hat{\mathbf{n}}^\ast$}

To maximise $\mathcal{F}(\hat{\mathbf{n}})$ subject to the unit-length constraint 
$\|\hat{\mathbf{n}}\|=1$, we introduce a Lagrange multiplier $\lambda$ and require
\begin{equation}
\label{eq:lagrange_condition}
\nabla_{\hat{\mathbf{n}}}\mathcal{F}(\hat{\mathbf{n}}) = \lambda\,\hat{\mathbf{n}}.
\end{equation}
Equivalently, the component of the gradient tangent to the unit sphere must vanish:
\begin{equation}
\label{eq:tangent_vanish}
\hat{\mathbf{n}} \times \nabla_{\hat{\mathbf{n}}}\mathcal{F}(\hat{\mathbf{n}}) = \mathbf{0}.
\end{equation}

To obtain $\nabla_{\hat{\mathbf{n}}}\mathcal{F}$, note that
\begin{equation}
\label{eq:grad_G_parallel}
\nabla_{\hat{\mathbf{n}}}G_{\parallel,hkl} = \nabla_{\hat{\mathbf{n}}}(\mathbf{G}_{hkl} \cdot \hat{\mathbf{n}}) = \mathbf{G}_{hkl},
\end{equation}
and, for $G_{\perp,hkl} \neq 0$,
\begin{equation}
\label{eq:grad_G_perp}
\nabla_{\hat{\mathbf{n}}}G_{\perp,hkl} = \nabla_{\hat{\mathbf{n}}}\sqrt{\|\mathbf{G}_{hkl}\|^2 - (\mathbf{G}_{hkl} \cdot \hat{\mathbf{n}})^2} 
= -\frac{G_{\parallel,hkl}}{G_{\perp,hkl}}\,\mathbf{G}_{hkl}.
\end{equation}
Hence,
\begin{align}
\nabla_{\hat{\mathbf{n}}}u_{hkl} &= R'\,\nabla_{\hat{\mathbf{n}}}G_{\perp,hkl} = -R'\frac{G_{\parallel,hkl}}{G_{\perp,hkl}}\mathbf{G}_{hkl}, \label{eq:grad_u} \\
\nabla_{\hat{\mathbf{n}}}v_{hkl} &= \frac{H'}{2}\,\nabla_{\hat{\mathbf{n}}}G_{\parallel,hkl} = \frac{H'}{2}\mathbf{G}_{hkl}. \label{eq:grad_v}
\end{align}

Applying the chain rule,
    \begin{equation}
\label{eq:chain_rule}
\nabla_{\hat{\mathbf{n}}} \bigl[f(u_{hkl})s(v_{hkl})\bigr] = \bigl[f'(u_{hkl})s(v_{hkl})\bigr]\nabla_{\hat{\mathbf{n}}}u_{hkl} 
+ \bigl[f(u_{hkl})s'(v_{hkl})\bigr]\nabla_{\hat{\mathbf{n}}}v_{hkl},
\end{equation}

we obtain
\begin{equation}
\label{eq:grad_F_explicit}
\begin{split}
\nabla_{\hat{\mathbf{n}}}\mathcal{F}(\hat{\mathbf{n}}) &= \sum_{h,k,l}^{\prime} \Biggl[ -R'\frac{G_{\parallel,hkl}}{G_{\perp,hkl}}\,f'(u_{hkl})\,s(v_{hkl}) + \frac{H'}{2}\,f(u_{hkl})\,s'(v_{hkl}) \Biggr] \mathbf{G}_{hkl}.
\end{split}
\end{equation}

A convenient explicit form for the scalar derivatives is
\begin{equation}
\label{eq:s_prime}
s'(v) = \frac{d}{dv}\left(\frac{\sin v}{v}\right) = \frac{v\cos v - \sin v}{v^2},
\end{equation}
and
\begin{equation}
\label{eq:f_prime}
f'(u) = \frac{d}{du}\left(\frac{2J_1(u)}{u}\right) = \frac{(J_0(u)-J_2(u))u - 2J_1(u)}{u^2}.
\end{equation}
(Any equivalent Bessel-identity form of $f'(u)$ is acceptable.)

Therefore, an optimal orientation $\hat{\mathbf{n}}^\ast$ must satisfy the (vector) stationarity equation
\begin{equation}
\label{eq:stationarity_full}
\begin{split}
\hat{\mathbf{n}}^\ast \times \sum_{h,k,l}^{\prime} \Biggl[ &-R'\frac{G_{\parallel,hkl}}{G_{\perp,hkl}}\,f'(u_{hkl})\,s(v_{hkl}) + \frac{H'}{2}\,f(u_{hkl})\,s'(v_{hkl}) \Biggr] \mathbf{G}_{hkl} = \mathbf{0},
\end{split}
\end{equation}
subject to $\|\hat{\mathbf{n}}^\ast\|=1$. Equivalently, defining
\begin{equation}
\label{eq:A_vector_def}
\begin{split}
\mathbf{A}(\hat{\mathbf{n}}) &= \sum_{h,k,l}^{\prime} \Biggl[ -R'\frac{G_{\parallel,hkl}}{G_{\perp,hkl}}\,f'(u_{hkl})\,s(v_{hkl})  + \frac{H'}{2}\,f(u_{hkl})\,s'(v_{hkl}) \Biggr] \mathbf{G}_{hkl},
\end{split}
\end{equation}
the extremum condition is simply that $\mathbf{A}(\hat{\mathbf{n}}^\ast)$ is parallel to $\hat{\mathbf{n}}^\ast$:
\begin{equation}
\hat{\mathbf{n}}^\ast \times \mathbf{A}(\hat{\mathbf{n}}^\ast) = \mathbf{0}.
\end{equation}

The expression is invariant under $\hat{\mathbf{n}} \to -\hat{\mathbf{n}}$ and under the cubic point group of the FCC lattice, so in numerical searches one may restrict $\hat{\mathbf{n}}$ to an irreducible wedge of the unit sphere.  In practice, one evaluates the reciprocal sum with a finite cutoff (e.g., $|h|,|k|,|l| \le n_{\max}$ or $\|\mathbf{G}\| \le G_{\max}$); the stationarity condition above then provides a smooth and readily computable criterion for candidate maxima. 
 The apparent factor $G_{\parallel}/G_{\perp}$ is not singular at $G_{\perp}=0$ because $f'(u) \sim -u/4$ as $u \to 0$, giving a finite limit for each term when $\hat{\mathbf{n}}$ aligns with a reciprocal vector.

\begin{figure}[t!]
    \centering
    \includegraphics[width=0.6\linewidth]{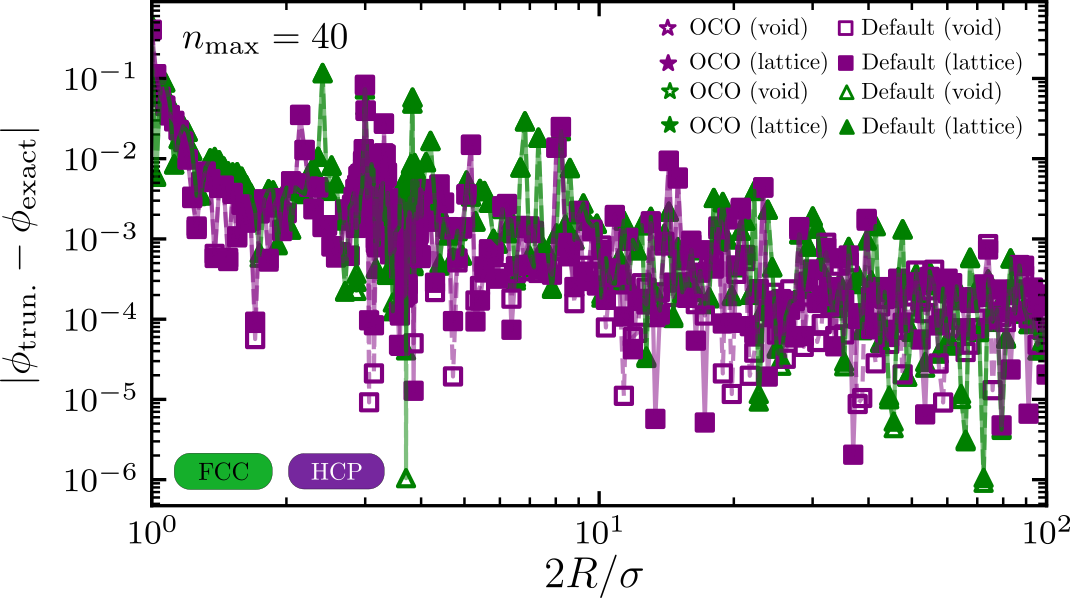}
    \caption{Example of the absolute difference between the truncated, \emph{i.e.,} $|h|,|k|,|l| \leq n_\text{max}$, reciprocal sum and the numerically exact packing fraction.}
    \label{fig:2_A1}
\end{figure}

%=========================================================
\section{Extension of Known Packings with a Cylindrical Sheath}

\begin{figure}
    \centering
    \includegraphics[width=0.8\linewidth]{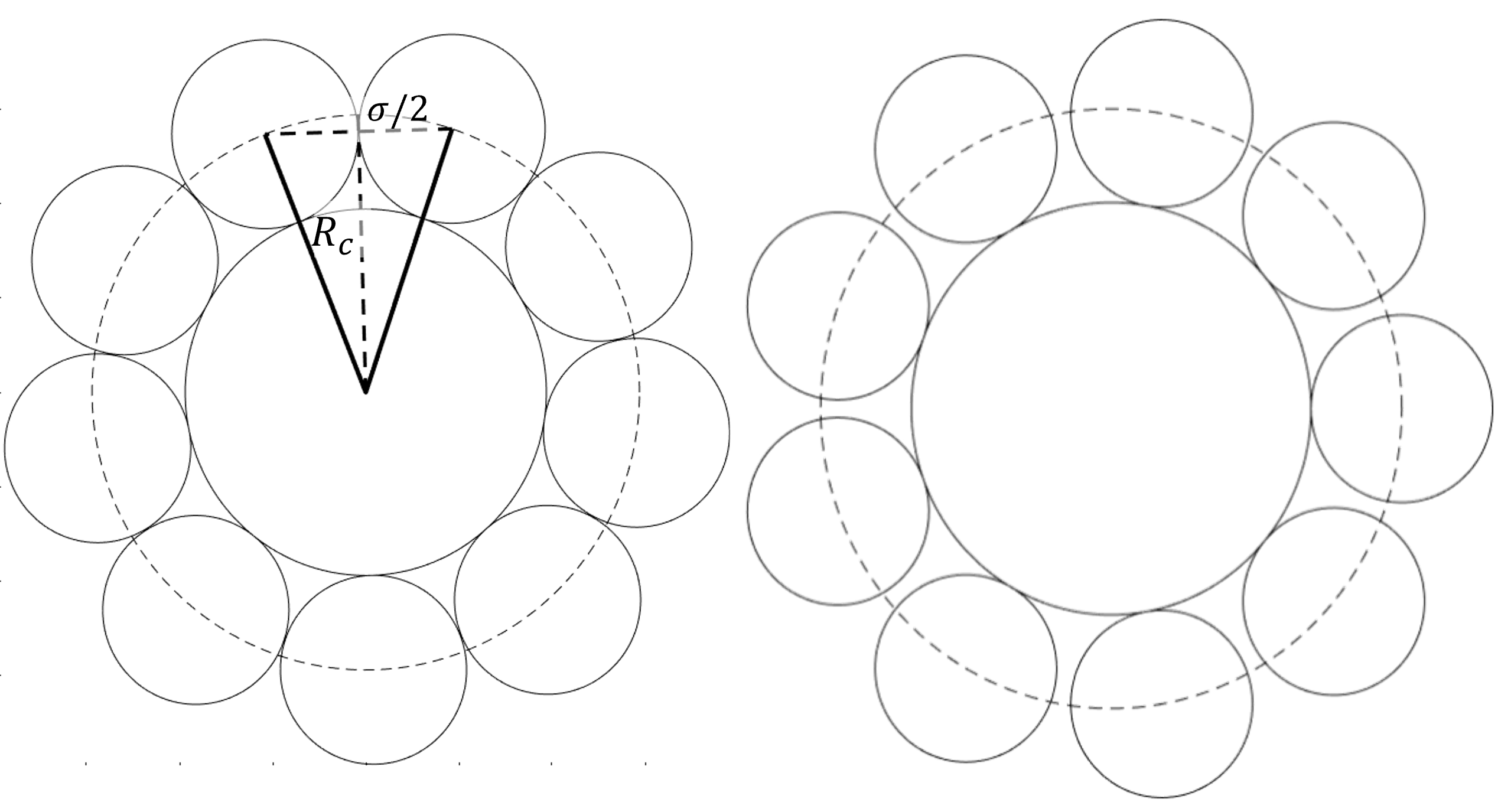}
    \caption{Planar arrangement of spheres around a central cylinder. The constraint between the radius and number assumes they are tightly packed, but they need not be.}
    \label{fig:s1}
\end{figure}

Consider a packing of spheres in a cylinder with a radius $R_0$ and a packing density $\phi_0$. In the simplest case, this may be a vertical column of spheres with $R_0=\sigma/2$ and $\phi_0=2/3$. In a given plane perpendicular to the axis of the cylinder, taken to be at $z=0$, a circular arrangement of $N_s$ spheres may be placed all tangent to the inner cylinder such that their centers all lie on a circle with radius $R_c=R_0+\sigma/2$ (Fig. \ref{fig:s1}). The maximum number of spheres that can lie on this circle is such that the distance between adjacent spheres is at least $\sigma$. An expression for $N_s$ arises from trigonometry:

\begin{equation}
    N_s=\bigg\lfloor \frac{\pi}{\arcsin({\sigma/2R_c})} \bigg\rfloor\approx \bigg\lfloor\frac{2\pi R_c}{\sigma}\bigg\rfloor.
\end{equation}

\begin{figure}
    \centering
    \includegraphics[width=0.8\linewidth]{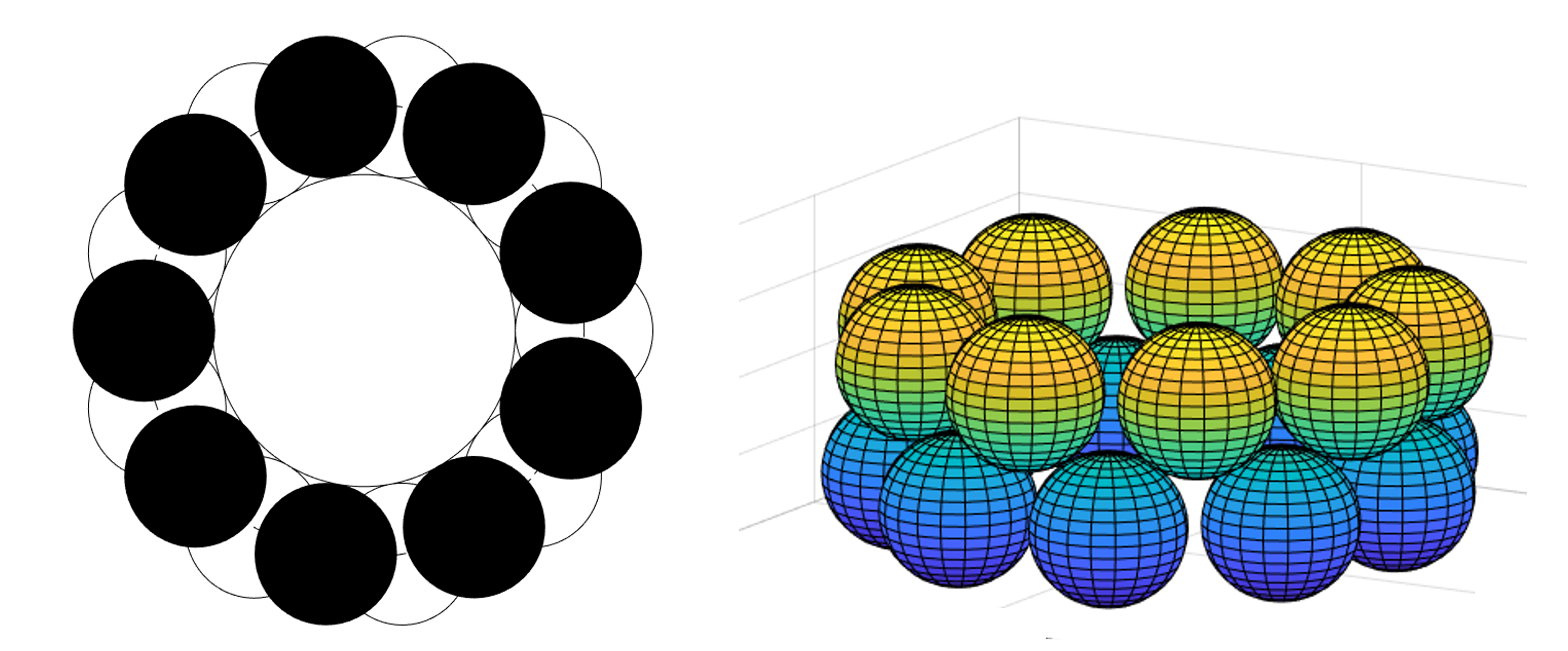}
    \caption{Arrangement of a second layer of spheres, in two and three dimensions.}
    \label{fig:s2}
\end{figure}

Although it is possible for the spheres on the circle to be tangent to their neighbors, this is generally not the case. A second ring of $N_s$ spheres may be placed on a similar circle raised a distance $\Delta z$ above the first. Since the angle subtended between two adjacent spheres is $2\pi/N_s$, the second ring of spheres may be rotated by half that angle, $\pi/N_s$, and the smallest value of $\Delta z$ allows each sphere to sit tangent to two below it (Fig. \ref{fig:s2}). $\Delta z$ may be determined by constraining the distance between a sphere at $\langle R_c,0,0\rangle$ and $\langle R_c\cos(\pi/N_s),R_c\sin(\pi/N_s),\Delta z \rangle$ to equal $\sigma$.

\begin{equation}
   \Delta z=\sigma\sqrt{1-2\frac{R_c^2}{\sigma^2}\left(1-\cos\frac{\pi}{N_s}\right)}
    \label{eq:xz}
\end{equation}

\begin{figure}
    \centering
    \includegraphics[width=0.8\linewidth]{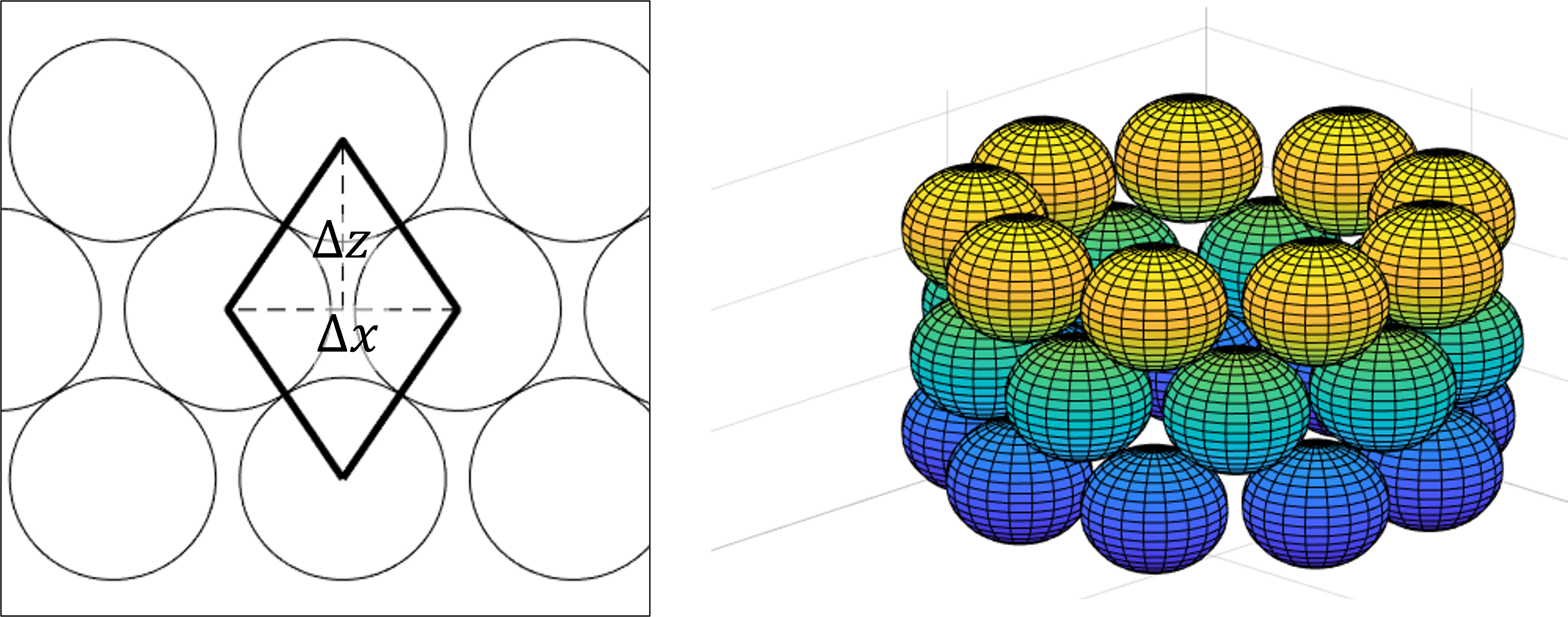}
    \caption{Arrangement of multiple layers of spheres into a cylindrical hexagonal packing. The area fraction of circles on the cylinder is determined by the area of the diamond, and the volume fraction of the cylindrical layer is two-thirds the area fraction of the cylinder.}
    \label{fig:s3}
\end{figure}

When the spheres are tightly packed in each ring, $\Delta z=\sqrt{3}\sigma/2$ as in a tight hexagonal packing, but $\Delta z$ is smaller when there is more distance between the spheres. More rings of spheres may be added with the same vertical offset and rotation, to produce a cylindrical layer of hexagonally packed spheres. The packing is not necessarily tight, as the spheres are only tangent to all their neighbors for certain values of $R_c$. To compute the volume fraction of this sheath, the cylinder that passes through the centers of all the spheres can be unrolled into a rectangle on which the circular cross-sections of the spheres form a planar hexagonal packing (Fig. \ref{fig:s3}). The centers of two horizontally adjacent circles, and the two circles tangent to both above and below, define a diamond with area $A_d$ that contains an entire circle. The area of this diamond is $A_d=\Delta x\Delta z$, and $\Delta x$ is defined along the surface of the cylinder, rather than through it.

\begin{equation}
    A_d=\Delta x\Delta z=\frac{2\pi R_c}{N_s}\sigma\sqrt{1-2\frac{R_c^2}{\sigma^2}\left(1-\cos\frac{\pi}{N_s}\right)}
\end{equation}

The area fraction $\phi_a$ is the area of a disk of diameter $\sigma$ divided by $A_d$:

\begin{equation}
\phi_a=\frac{\pi(\sigma/2)^2}{A_d}=\frac{N_s\sigma}{8R_c\sqrt{1-2\frac{R_c^2}{\sigma^2}\left(1-\cos\frac{\pi}{N_s}\right)}}   
\end{equation}

This cylindrical monolayer of spheres lies in an annulus of thickness $\sigma$ with inner radius $R_0$ and outer radius $R_E=R_0+\sigma$. Since each sphere in this annulus may be thought of as occupying its own sub-cylinder of diameter and height $\sigma$ (with an axis along the radial vector between center of each sphere and the central axis of the entire cylinder), the volume fraction of the annular sheath $\phi_s$ is 2/3 the area fraction $\phi_a$ of the unwrapped cylinder (2/3 being the volume fraction of a cylinder bounding a sphere). The annular sheaths have the highest packing fraction when the spheres are tangent to their neighbors, which occurs at the smallest radius that allows a new sphere to be added. Generally speaking, this occurs in intervals of $\Delta R_c=1/\pi$ (in units of $\sigma=2$), beginning with six spheres at $R_c=\sigma$.

Now, we consider the sphere packing fraction of a cylinder of radius $R_E=R_0+\sigma$ and height $H$, consisting of an inner cylinder with packing fraction $\phi_0$ and an annular sheath with packing fraction $\phi_s$. This derivation assumes that $H\gg R_0$, such that potential issues of incommensurability between the tops of spheres in the central cylinder and in the annulus do not significantly affect the packing fraction. We first consider the total volume of the spheres in both these components, which is the number of spheres times the volume of a single sphere $\pi \sigma^3/6$.

\begin{equation}
V_{spheres}=(\pi R_0^2 H)\phi_0 +(\frac{2}{3}\phi_a) (2\pi\sigma R_c H)
\end{equation}

The volume fraction $\phi'$ may be found by dividing this by the entire volume of the new cylinder.
\begin{equation}
\phi'=\frac{V_{spheres}}{\pi R_E^2H}=\frac{1}{R_E^2}\left(\phi_0R_0^2+\frac{N_s\sigma^2}{6\sqrt{1-2\frac{R_c^2}{\sigma^2}\left(1-\cos\frac{\pi}{N_s}\right)}}   \right)  
\label{eq:phiprime}
\end{equation}

The full expression may be written in terms of $R_0$ and $\phi_0$:

\begin{equation}
\phi'=\frac{1}{(R_0+\sigma)^2}\left(\phi_0R_0^2+\frac{\lfloor \frac{\pi}{\arcsin(\frac{\sigma}{2R_0+\sigma})} \rfloor \sigma^2}{6\sqrt{1-2\frac{(R_0+\sigma/2)^2}{\sigma^2} \left( 1 -\cos(\frac{\pi}{\lfloor \frac{\pi}{\arcsin(\frac{\sigma}{2R_0+\sigma})} \rfloor }) \right)}}   \right)   
\end{equation}

This process can be repeated recursively. After the first sheath is added, the combined core-sheath cylinder may serve as the core cylindrical packing for the addition of a second sheath. The fixed point of the packing fraction is the volume fraction of a large cylindrical monolayer, which is $(2/3)(\pi/\sqrt{12})=\pi/\sqrt{27}\approx0.604$. This is not competitive with the OCO beyond $2R/\sigma\approx 8$, but yields denser packing in the $4 \lessapprox 2R/\sigma \lessapprox 8$ range.

\end{document}